\documentclass[preprint,12pt]{elsarticle}

\usepackage{lineno,hyperref}
\usepackage{multirow}
\usepackage{graphicx}
\usepackage{graphics}
\usepackage{epstopdf}
\usepackage{xcolor}

\usepackage{url}
  \usepackage{breakurl}
\usepackage{mathrsfs}
\usepackage{subfigure}
\usepackage{amssymb}
\usepackage{amsmath}
\usepackage{color}
\usepackage{ulem}
\numberwithin{equation}{section}
\allowdisplaybreaks

\newtheorem{theorem}{Theorem}[section]
\newtheorem{lemma}{Lemma}[section]

\modulolinenumbers[5]
\usepackage{amssymb}
\begin{document}

\begin{frontmatter}



\title{A Data-Driven Network Model for the Emerging COVID-19 Epidemics in Wuhan, Toronto and Italy}


%\author{}
%\cortext[mycorrespondingauthor]{Corresponding author, Contributed equally}
%\ead{support@elsevier.com, lxue@hrbeu.edu.cn}
%\fntext[firstauthor]{Contributed equally}

%\author[address1]{Ling Xue \fnref{firstauthor}}
\author{{Ling Xue}$^{\text{a},\#}$\footnote{$\#$ Contributed equally}}
\author[address1]{Shuanglin Jing}
\author[address2]{Joel C. Miller}
\author{{Wei Sun}$^{\text{a},\#}$}
\author[address1]{Huafeng Li}
\author[address3]{Jos\'e  Guillermo Estrada-Franco\corref{mycorrespondingauthor}}
\ead{jestradaf@ipn.mx}
\author[address4]{James M Hyman}
\author[address5]{Huaiping Zhu\corref{mycorrespondingauthor}}
\ead{huaiping@mathstat.yorku.ca}
%\address{}
\address[address1]{College of Mathematical Sciences, Harbin Engineering University, Harbin, Heilongjiang, 150001, China}
\address[address2]{School of Engineering and Mathematical Sciences, Melbourne, La Trobe University, 3086, Australia}
\address[address3]{Instituto Polit\'ecnico Nacional, Centro de Biotecnolog\'ia Gen\'omica, Cd. Reynosa, Tamaulipas, 88710, M\'exico}
\address[address4]{Department of Mathematics, Tulane University, New Orleans, LA, 70118, USA}
\address[address5]{Lamps and Center of Disease Modelling (CDM), Department of Mathematics and Statistics, York University, Toronto, ON, M3J 1P3, Canada}

\cortext[mycorrespondingauthor]{Corresponding author, Contributed equally}

\begin{abstract}
The ongoing Coronavirus Disease 2019 (COVID-19) pandemic threatens the health of humans and causes great economic losses. Predictive modelling and forecasting the epidemic trends are essential for developing countermeasures to mitigate this pandemic.  We develop a network model, where each node represents an individual and the edges represent contacts between individuals where the infection can spread. The individuals are classified based on the number of contacts they have each day (their node degrees) and their infection status. The transmission network model was respectively  fitted to the reported data for the COVID-19 epidemic in Wuhan (China), Toronto (Canada), and the Italian Republic using a Markov Chain Monte Carlo (MCMC) optimization algorithm.  Our model fits all three regions well with narrow confidence intervals and could be adapted to simulate other megacities or regions. The model projections on the role of containment strategies can help inform public health authorities to plan control measures.
%The ongoing outbreak of the 2019 novel Coronavirus Disease (COVID-19) has caused
%global pandemic, threatening the health of humans and causing great economic losses.
%Since the reported cases of coronavirus disease (COVID) are rising worldwide, forecasting
%the epidemic trends and designing effective mitigation strategies are essential for mitigating the epidemic spread with limited resources. In this study, we propose and analyze
%a network model, taking into account heterogeneity to predict the trends and mitigate
%the spread of COVID-19 epidemics. The basic reproduction number and final size of
%infection are derived analytically. The model was applied to predict the epidemic trends for Wuhan, Italy, and Toronto, and various mitigation strategies have been test on
%the framework. The findings may provide guidance for public health authorities to plan
%control measures in advance.
\end{abstract}




\end{frontmatter}


%\section{}
%
%\label{}
\section{Introduction}

The development of international trade and tourism has accelerated the spatial spread of infectious diseases.  The limited data available on emerging epidemics adds to the challenge of mitigating the spread of emerging infections \cite{anderson1991}.
The unprecedented Coronavirus Disease 2019 (COVID-19) outbreak began at the end of 2019.  The number of reported cases keeps rising worldwide and thousands of lives have been claimed. This pandemic is having an enormous impact on world health, disturbing the stability of the societies, and triggers great economic losses. Predicting the future of the pandemic, assessing the impacts of current interventions, and evaluating the effectiveness of alternate mitigation strategies are of utmost importance for saving lives.

Mathematical models can be used to understand the dynamics of epidemics and help %formulate
inform control strategies. A numerous number of  %wide variety of
models are being used to project the current COVID-19 pandemic. Ziff and Ziff analyzed the number of reported cases for Wuhan (China) and showed that the growth of the daily number of confirmed new cases indicates an underlying fractal or small-world network of connections between susceptible and infected individuals \cite{ziff2020fractal}.
Wang et al. developed an SEIR model to estimate the epidemic trends in Wuhan, assuming the prevention and control measures were either sufficient or insufficient to control the epidemic \cite{wang2020phase}.
Kucharski et al. combined a stochastic transmission model with data on cases of COVID-19 in Wuhan and international cases to estimate how the transmission had varied over time between January and February in 2020 \cite{kucharski2020early}.

Kraemer et al. analyzed the impact of interventions on the spread of COVID-19 in China using transportation data \cite{kraemer2020effect}.
Chinazzi et al. used a global meta-population disease transmission model to project the impact of travel limitations on the national and international spread of the epidemic.  They showed that the travel restriction of Wuhan, China had a more marked effect on the international scale than that on Mainland China \cite{chinazzi2020effect}.

Ferguson et al. found that optimal mitigation policies (combining home isolation of suspected cases, home quarantine of those living in the same household as suspected cases, and social distancing of the elderly and others at most risk of severe disease) might reduce peak healthcare demand by 2/3 and deaths by half \cite{Ferguson_2020}.
Likewise, Hellewell et al. developed a stochastic transmission model and found that highly effective contact tracing and case isolation is enough to control a new outbreak of COVID-19 within three months in most scenarios \cite{hellewell2020feasibility}.

Zhang et al. fitted the reported serial interval (mean and standard deviation) with a gamma distribution to estimate the basic reproduction number at the early stage of a COVID-19 outbreak, indicating the potential of second outbreaks \cite{Zhang2020}.
Maier et al. developed a compartmental model dividing individuals into susceptible, exposed, removed, and quarantined symptomatically infected and showed that the distinctive subexponential increase of confirmed cases in mainland China could be explained as a direct consequence of containment policies that effectively deplete the susceptible population \cite{maier2020effective}.

Most of these models are based on assuming the population is homogeneously mixing,  that is, the contacts between people are random and uniformly distributed throughout the population. However, different individuals may have varying numbers of acquaintances and contacts in the real world.  The important role that heterogenous contact networks play in the transmission dynamics of infectious diseases is often overlooked \cite{welch2011statistical}.
Models that take into account contact heterogeneity better represent the actual transmission network through a population and are more likely to capture the true epidemic dynamics.

Disease propagation is closely linked with the structure of social contact networks \cite{moore2000epidemics}. The ubiquitous diversity in contact patterns and heterogeneity among individuals depends on differences in social structures, spatial distances, and behavior \cite{bansal2007individual}.
The heterogeneity exists at a wide range of scales and leads to highly variable transmission dynamics of infectious diseases \cite{hochberg1991non, addy1991generalized}.

Many real-world social networks can be characterized by a random Watts-Strogatz (WS) small-world network \cite{watts1998collective, boccaletti2006complex}.
In a small-world network, most nodes can be reached from every other node by a small number of hops or steps, even if they are not immediate neighbours.  This type of network model allows us to adapt changes to some realistic network structures and examine the effects of control and intervention countermeasures  such as social distancing, self-isolation, and personal protection. The framework and analysis can be applied to study the transmission dynamics in different regions and many other infectious diseases.

The COVID-19 epidemic in Wuhan ended in April, while the epidemics in the Greater Toronto Area (GTA, Canada) and the Italian Republic are continuing to grow. We fit the parameters of our network model to the confirmed cases in each of these regions.
Although Wuhan, Toronto, and Italy differ in some ways,  the way that SARS-CoV-2 is transmitted from one person to another is quite similar. Individuals may acquire infection from other infectious individuals, even if they do not  contact each other directly.
The Watts-Strogatz model supplies an ideal tool 
to study the spread of epidemics among individuals even if  their locations are not considered. 
We used the Watts-Strogatz model to generate random networks with the small world properties appropriate for infectious disease transmission in these cities
\cite{watts1998collective,  boccaletti2006complex}.

The epidemic curves are all fitted very well using the small-world network structure models, indicating that the typical small-world property is able to capture the contact patterns during COVID-19 epidemics. The differences in these fitted parameters and starting times reflect the differences in the underlying transmission mechanisms and potential spread in the regions. The model then projected the trends of COVID-19 spread by simulating epidemics in the Wuhan, Toronto, and Italy networks. Our findings can guide public health authorities to implement effective mitigation strategies and be prepared for potential future outbreaks.

\section{The network  model}  \label{Model}
We develop a network-based model by extending the network SIR model \cite{boguna2003statistical}
by incorporating the characteristics of COVID-19 transmission to assess the spread of the disease in heterogeneous populations. We derive the explicit expression of the epidemic threshold and discuss the  final epidemic size for the network model.

\subsection{Model formulation}
We classify individuals by their average number of contacts in a typical day (time unit for the modeling) represented on the network by their degree
$k\  (k=1, 2, \cdots, n)$. Individuals with degree $k$ are divided into  susceptible ($S_k$), exposed ($E_k$), asymptomatically infected ($A_k$), symptomatically infected ($I_k$), hospitalized  ($H_k$), recovered  ($R_k$), and dead  ($D_k$) states.  Our model is formulated as follows
\begin{eqnarray}{\label{E:mainmodel}}
\left\{
\begin{array}{lll}
\frac{ ds_k}{dt}&=- \beta k s_k\sum_{k'}[1-(1-\frac{k'-1}{k'}P(k'|k)i_{k'})(1-\frac{k'-1}{k'}P(k'|k)\sigma a_{k'})], \\ \vspace{3pt}
\frac{ de_k}{dt}&=\beta k s_k\sum_{k'}[1-(1-\frac{k'-1}{k'}P(k'|k)i_{k'})(1-\frac{k'-1}{k'}P(k'|k)\sigma a_{k'})] -\epsilon e_k,  \\ \vspace{3pt}
\frac{da_k}{dt}&=(1-\delta)\epsilon e _k-\gamma_a a_k,    \\ \vspace{3pt}
\frac{ di_k}{dt}&=\delta\epsilon e _k -\gamma i_k-\mu i_k-\xi i_k , \\ \vspace{3pt}
\frac{ dh_k}{dt}&=\xi i_k-\gamma_{h} h_k-\mu h_k, \\ \vspace{3pt}
\frac{dr_k}{dt}&=\gamma i_k+\gamma_a a_k+\gamma_{h} h_k, \\ \vspace{3pt}
\frac{dd_k}{dt}&=\mu i_k+\mu h_k,
\end{array}
\right.
\end{eqnarray}
where $s_k=S_k/N_k$,  $e_k=E_k/N_k$, $a_k=A_k/N_k$, $i_k=I_k/N_k$, $h_k=H_k/N_k$, $r_k=R_k/N_k$, and $d_k=D_k/N_k$ represent the fractions of susceptible, exposed, asymptomatically infected, symptomatically infected, hospitalized, recovered, and dead individuals with degree $k$, respectively.
Here, $N_k$ is the total number of individuals with degree $k$, and $N_k=$ $S_k$ $+E_k$ $+A_k+I_k$  $+H_k+R_k$ $+D_k$, and
$s_k$ $+e_k$ $+a_k+i_k$  $+h_k+r_k$ $+d_k=1$.
%Here, $N_k$ is the total number of individuals with degree $k$,
$P(k'|k)$ represents the probability that an edge from a node with degree $k$ connects to a node with degree $k'$.
For uncorrelated networks, $P(k'|k) =k'P(k')/ \langle k \rangle$ \cite{Barrat2008}.  Since the node with degree $k'$ shares an edge with the node degree $k$, and only has $(k'-1)$ free edges, a fraction $\frac{k'-1}{k'}$  of nodes may acquire the infection.

We assume that the transmission rates of symptomatically infected individuals and asymptomatically infected individuals are $\beta$ and $\sigma\beta$, respectively. The factor $\sigma$ accounts for the different transmission rates between asymptomatically infected individuals and symptomatically infected individuals. $ \beta k s_k\sum_{k'}\frac{k'-1}{k'}P(k'|k)i_{k'}$ represents the fraction of nodes with degree $k$   infected by symptomatically infected nodes, and  $ \sigma\beta k s_k\sum_{k'}\frac{k'-1}{k'}P(k'|k)a_{k'}$ $=$ $ \beta k s_k\sum_{k'}\frac{k'-1}{k'}P(k'|k)\sigma a_{k'}$ represents the fraction of nodes with degree $k$  infected by asymptomatically infected nodes. Here, $\frac{k'-1}{k'}P(k'|k)i_{k'}$ represents the probability that an edge from a degree $k$ node connects to a symptomatically infected node with degree $k'$, and $\frac{k'-1}{k'}P(k'|k)\sigma a_{k'}$ represents the probability that an edge from a degree $k$ node connects to an asymptomatically infected node with degree $k'$.

%Hence,
In Model \eqref{E:mainmodel}, the term
 $(1-\frac{k'-1}{k'}P(k'|k)i_{k'})$ represents the probability of not being infected by a symptomatically infected node with degree $k'$, and $(1-\frac{k'-1}{k'}P(k'|k)\sigma a_{k'})$  represents the probability of not being infected by an asymptomatically infected node with degree $k'$. Thus, $(1-\frac{k'-1}{k'}P(k'|k)i_{k'})(1-\frac{k'-1}{k'}P(k'|k)\sigma a_{k'})$ is the probability that a  node will neither  be infected by a symptomatically infected  nor be infected by an asymptomatically infected neighbor with degree $k'$, and $1-(1-\frac{k'-1}{k'}P(k'|k)i_{k'})(1-\frac{k'-1}{k'}P(k'|k)\sigma a_{k'})$ is the probability of being infected by a symptomatically infected  or  an asymptomatically infected neighbor with degree $k'$.

Therefore, the susceptible individuals are infected at   rate
$$ \beta k s_k\sum_{k'}[1-(1-\frac{k'-1}{k'}P(k'|k)i_{k'})(1-\frac{k'-1}{k'}P(k'|k)\sigma a_{k'})]$$
and enter the exposed state.
After incubation period with a mean time of $1/\epsilon$ days,
exposed individuals become symptomatically infected and asymptomatically infected with probabilities $\delta$ and $1-\delta$, respectively. Symptomatically infected individuals are hospitalized at rate $\xi$, and die at rate $\mu$. Asymptomatically infected individuals, symptomatically infected individuals, and hospitalized individuals recover at rates $\gamma_a$, $\gamma$, and $\gamma_h$, respectively. Both the hospitalized individuals and  symptomatically infected individuals die at rate $\mu$.

\subsection{Mathematical analysis}

We derive the epidemic threshold to predict whether the epidemic will spread or die out and derive final epidemic size  to quantify the total number of infected individuals.
\subsubsection {The epidemic threshold}
To estimate the transmission potential of the epidemic, we derive the important epidemic threshold, $R_0$, defined as the average number of secondary cases produced by an infected individual in a completely susceptible population \cite{diekmann1990definition}.
There exists a  disease-free equilibrium,
\begin{equation*}
\begin{split}
&(s_1,\cdots, s_n, e_1,\cdots, e_n, a_1,\cdots, a_n, i_1,\cdots, i_n, h_1,\cdots, h_n, r_1,\cdots, r_n, d_1,\cdots, d_n)^T\\
&=(1, \cdots , 1, 0,\cdots , 0, 0,\cdots , 0,0,\cdots , 0,0,\cdots , 0,0,\cdots , 0, 0, \cdots, 0)^T =: E_0.
\end{split}
\end{equation*}
We compute $R_0$ following the next generation matrix approach presented by van den Driessche and Watmough \cite{van}. For simplicity, we only consider the compartments related to infection, namely, $e_k$, $a_k$ and $i_k$, and rewrite the equations as the difference between vectors  $\mathcal{F}_k$ and $\mathcal{V}_k$   following the notations in \cite{van}
 $$[ \frac{de_k}{dt}, \frac{da_k}{dt}, \frac{di_k}{dt}]^T=\mathcal{F}_k- \mathcal{V}_k,$$
 where
 \[
  \mathcal{F}_k= (\mathcal{F}_{ik})=\left[\begin{array}{c}
  \beta k s_k\sum_{k'}[1-(1-\frac{k'-1}{k'}P(k'|k)i_{k'})(1-\frac{k'-1}{k'}P(k'|k)\sigma a_{k'})]\\
  0\\
  0\\
  \end{array}\right],
\]
\[
  \mathcal{V}_k= (\mathcal{V}_{ik})=\left[\begin{array}{c}
 \epsilon e_k\\
  -(1-\delta)\epsilon e_k +\gamma_a a_k\\
 -\delta \epsilon e_k +\gamma i_k+\mu i_k+\xi i_k\\
  \end{array}\right].
\]
Here, $\mathcal{F}_{ik}$ represents the rate at which new infections are produced and $\mathcal{V}_{ik}$ represents the rate at which individuals transfer between compartments, $i=1, 2, 3$ and $k=1, \cdots, n$ for Model \eqref{E:mainmodel}.

The Jacobian matrix $F$ is
\begin{equation*}
F=\left[\frac{\partial \mathcal{F}_{ik} }{\partial z_j}\right]_{E_0}=\left[\begin{array}{cccc}
 0_{n \times n}  & \sigma\beta F' & \beta F' \\
0_{n \times n}&0_{n \times n} &0_{n \times n}\\
0_{n \times n}&0_{n \times n} &0_{n \times n}\\
  \end{array}\right],
\end{equation*}
 where $z= (z_j) = (e_1, \cdots, e_n, a_1, \cdots, a_n, i_1, \cdots, i_n)$ and
\begin{equation*}
\begin{split}
F'&=\left[\begin{array}{cccc}
0  & \frac{1}{2}P(2|1)& \cdots&\frac{n-1}{n}P( n|1) \\
0  & P( 2|2) & \cdots& \frac{2(n-1)}{n}P( n|2) \\
\vdots&  \vdots& \vdots &  \vdots \\
0&\frac{n}{2}P( 2|n) & \cdots &(n-1)P( n|n) \\
  \end{array}\right]\\
&=\frac{1}{\langle k\rangle }\left[\begin{array}{cccc}
0  & P(2)& \cdots&(n-1)P(n) \\
0  & 2P(2) & \cdots& 2(n-1)P(n) \\
\vdots&  \vdots& \vdots &  \vdots \\
0&nP(2) & \cdots &n(n-1)P(n) \\
  \end{array}\right].
\end{split}
\end{equation*}

The matrices $V$ and $V^{-1}$ are
%\footnotesize{
\begin{equation*}
V=\left[\frac{\partial \mathcal{V}_i }{\partial z_j}\right]_{E_0}=\begin{bmatrix}
  \epsilon \mathbf I_n &0_{n\times n} &0_{n\times n}\vspace{1pt}\\
  -(1-\delta)\epsilon \mathbf I_n & \gamma_{a} \mathbf I_n & 0_{n\times n}\vspace{1pt}\\
  -\delta\epsilon \mathbf I_n & 0_{n\times n} & (\gamma+\mu+\xi) \mathbf I_n
\end{bmatrix},
 \end{equation*}
% }
where $\mathbf I_n$ is the $n\times n$ identity matrix, and
\begin{equation*}
\setlength{\arraycolsep}{2pt}
V^{-1}= \begin{bmatrix}
  \frac{1}{\epsilon}\mathbf I_n & 0_{n\times n}& 0_{n\times n}\vspace{1pt}\\
  \frac{1-\delta}{\gamma_{a}} \mathbf I_n & \frac{1}{\gamma_a} \mathbf I_n & 0_{n\times n}\vspace{1pt}\\
  \frac{\delta}{ \gamma +\mu+\xi }\mathbf I_n & 0_{n\times n} & \frac{1}{\gamma +\mu+\xi}\mathbf I_n
\end{bmatrix}.
%\left[ \begin{array}{cccccccccccc}
%	\frac{1}{\epsilon}&		0&		\cdots&		0&		0&		0&		\cdots&		0&		0&		0&		0&		0\\
%	0&		\frac{1}{\epsilon}&		\cdots&		0&		0&		0&		\cdots&		0&		0&		0&		0&		0\\
%	\vdots&		\vdots&		\ddots&		\vdots&		\vdots&		\vdots&		\cdots&		\vdots&		\vdots&		\vdots&		\vdots&		\vdots\\
%	0&		0&		\cdots&		\frac{1}{\epsilon}&		0&		0&		\cdots&		0&		0&		0&		0&		0\\
%	\frac{1-\delta}{\gamma_{a}}&		0&		\cdots&		0&		\frac{1}{\gamma_a}&		0&		\cdots&		0&		0&		 0&		0&		 0\\
%	0&			\frac{1-\delta}{\gamma_{a}}&		\cdots&		0&		0&		\frac{1}{\gamma_a}&		\cdots&		0&		0&		 0&		0&		 0\\
%	\vdots&		\vdots&		\ddots&		\vdots&		\vdots&		\vdots&		\ddots&		\vdots&		\vdots&		\vdots&		\vdots&		\vdots\\
%	0&		0&		\cdots&			\frac{1-\delta}{\gamma_{a}}&		0&		0&		\cdots&		\frac{1}{\gamma_a}&		0&		0&		 \cdots&		 0\\
%	\frac{\delta}{ \gamma +\mu+\xi }&		0&		\cdots&		0&		0&		0&		\cdots&		0&		\frac{1}{\gamma +\mu+\xi}&		0&		 \cdots&		 0\\
%	0&		\frac{\delta}{ \gamma +\mu+\xi }&		\cdots&		0&		0&		0&		\cdots&		0&		0&		\frac{1}{\gamma +\mu+\xi}&		 \cdots&		 0\\
%	\vdots&		\vdots&		\cdots&		\vdots&		\vdots&		\vdots&		\ddots&		\vdots&		\vdots&		\vdots&		\ddots&		\vdots\\
%	0&		0&		\cdots&		\frac{\delta}{ \gamma +\mu+\xi }&		0&		0&		\cdots&		 0&		0&		0&		\cdots&		\frac{1}{\gamma +\mu+\xi}\\
%\end{array} \right].
\end{equation*}
The next generation matrix is
\begin{equation}\label{ngmatrix}
FV^{-1}=\left[\begin{array}{cccc}
\beta  \left(\frac{1-\delta}{\gamma_a}\sigma+\frac{\delta}{\gamma +\mu+\xi}\right)F'  &\frac{\sigma\beta }{\gamma_a} F' & \frac{ \beta }{\gamma +\mu+\xi}F' \vspace{1pt}\\
0_{n \times n}&0_{n \times n} &0_{n \times n}\vspace{1pt}\\
0_{n \times n}&0_{n \times n} &0_{n \times n}
  \end{array}\right].
\end{equation}
Since the rank of matrix $F'$ is  $1$, the spectral radius of $F'$ is its trace, i.e.,
\begin{align*}
\rho(F')={\rm Tr}(F')=\frac{1}{\langle k \rangle}\sum_{k} (k-1)kP(k)
=\frac{\langle k^2 \rangle-\langle k \rangle}{\langle k \rangle}.
\end{align*}

It follows from %By
\eqref{ngmatrix} that the basic reproduction number $R_0$ becomes
\begin{equation*}
  \begin{split}
R_0= \rho(FV^{-1}) %&=\beta \left(\frac{1-\delta}{\gamma_a}\sigma+\frac{\delta}{\gamma +\mu+\xi}\right) \rho(F')\\
   &=\beta  \left(\frac{1-\delta}{\gamma_a}\sigma+\frac{\delta}{\gamma +\mu+\xi}\right) \frac{(\langle k^2 \rangle-\langle k \rangle)}{\langle k \rangle},
  \end{split}
\end{equation*}
where $ \beta  \frac{1-\delta}{\gamma_a}\sigma$ and $ \beta\frac{\delta}{\gamma +\mu+\xi}$ represent the average numbers of secondary cases produced by an asymptomatically infected individual and a symptomatically infected individual in a homogeneously mixed population, respectively.
The term $\frac{\langle k^2 \rangle-\langle k \rangle}{\langle k \rangle}$
represents the average excess degree of nodes in the network \cite{wang2015revisiting}.

\subsubsection{Final epidemic size}
We shall derive the final size following the approach in \cite{wang2019further}. The nonlinear term in the first and second equations of  Model (2.1) can be rewritten as
 \begin{align*}
& \beta k s_k\sum_{k'}[1-(1-\frac{k'-1}{k'}P(k'|k)i_{k'})(1-\frac{k'-1}{k'}P(k'|k)\sigma a_{k'})]\\
 &=\beta k s_k\sum_{k'}[1-1+\frac{k'-1}{k'}P(k'|k)i_{k'}+\frac{k'-1}{k'}P(k'|k)\sigma a_{k'}-(\frac{k'-1}{k'}P(k'|k))^2i_{k'}\sigma a_{k'}]\\
 &=\beta k s_k\sum_{k'}[\frac{k'-1}{k'}P(k'|k)i_{k'}+\frac{k'-1}{k'}P(k'|k)\sigma a_{k'}-(\frac{k'-1}{k'}P(k'|k))^2i_{k'}\sigma a_{k'}].
  \end{align*}
  When $i_{k'}\ll 1$ and $a_{k'}\ll 1$,  $i_{k'}a_{k'} \approx 0 $. Hence,
   \begin{align*}
& \beta k s_k\sum_{k'}[1-(1-\frac{k'-1}{k'}P(k'|k)i_{k'})(1-\frac{k'-1}{k'}P(k'|k)\sigma a_{k'})]\\
 &\approx \beta k s_k\sum_{k'}[\frac{k'-1}{k'}P(k'|k)(i_{k'}+\sigma a_{k'})].
  \end{align*}
Hence, Model \eqref{E:mainmodel} can be simplified as
\begin{eqnarray}{\label{E:simp_model}}
\left\{
\begin{array}{lll}
\frac{ d s_k}{dt}&=- \beta k s_k\sum_{k'}\frac{k'-1}{k'}P(k'|k)\left(i_{k'}+\sigma a_{k'}\right). \vspace{3pt}\\
\frac{ d e_k}{dt}&=\beta k s_k\sum_{k'}\frac{k'-1}{k'}P(k'|k)\left(i_{k'}+\sigma a_{k'}\right) -\epsilon e_k,\vspace{3pt}\\
%\frac{d a_k}{dt}&=\epsilon e _k-\nu a_k -\gamma^a a_k, \\%  \label{A'}\vspace{3pt} \\
%\frac{ d i_k}{dt}&=\nu a_k-\gamma i_k-\mu i_k,\\% \label{I'}\vspace{3pt}\\
%\frac{d r_k}{dt}&=\gamma i_k+\gamma^a a_k, \\%\label{R'}\vspace{3pt}\\
%\frac{d d_k}{dt}&=\mu i_k.\vspace{3pt}\\
\frac{da_k}{dt}&=(1-\delta)\epsilon e _k-\gamma_a a_k, \vspace{3pt}   \\
\frac{ di_k}{dt}&=\delta\epsilon e _k -\gamma i_k-\mu i_k-\xi i_k , \vspace{3pt}\\
\frac{ dh_k}{dt}&=\xi i_k-\gamma_{h} h_k-\mu h_k,\vspace{3pt} \\
\frac{dr_k}{dt}&=\gamma i_k+\gamma_a a_k+\gamma_{h} h_k, \vspace{3pt}\\
\frac{dd_k}{dt}&=\mu i_k+\mu h_k.
\end{array}
\right.
\end{eqnarray}

We first claim that $\lim_{t\to +\infty}e_k(t)=0$.
Otherwise, there exist $T, \eta >0$ such that $e_k(t)\ge \eta > 0$, $\forall t>T$ since $e_k(t)\geq 0$.
Since $\frac{ d }{dt}\left(s_k+  e_k\right)= -\epsilon e_k \le -\epsilon \eta$,
$s_k(t)+  e_k(t) \le -\epsilon \eta t + s(T)+e(T)$.
Therefore, $\lim_{t \to \infty}(s_k(t)+  e_k(t)) = -\infty$, leading to a contradiction.
Similarly, we can show that
 $${\lim_{t\to +\infty}e_k(t)=\lim_{t\to{+\infty}}a_k(t)=\lim_{t\to{+\infty}}i_k(t)=\lim_{t\to{+\infty}}h_k(t)=0},\quad \forall k.$$
%In Model (\ref{E:simp_model}), $\frac{ d s_k}{dt}\le 0 $, and
%$$\frac{ d }{dt}\left(s_k+  e_k\right)= -\epsilon e_k \le 0, $$
%$$\frac{ d }{dt}\left(s_k+  e_k+a_k\right)=-\left(\delta\epsilon e _k+\gamma_a a_k \right) \le 0,$$
%$$\frac{ d }{dt}\left(s_k+  e_k+a_k+i_k\right)=-\gamma_a a_k-\left(\gamma +\mu +\xi\right) i_k \le 0,$$
%$$\frac{ d }{dt}\left(s_k+  e_k+a_k+i_k+h_k\right)=-\left(\gamma_{h}+\mu \right)h_k-\left(\gamma +\mu \right) i_k \le 0,$$
%which imply that $${\lim_{t\to +\infty}e_k(t)=\lim_{t\to{+\infty}}a_k(t)=\lim_{t\to{+\infty}}i_k(t)=\lim_{t\to{+\infty}}h_k(t)=0}$$ for all $k$.
%Hence, the disease will eventually die out.

For a homogeneous network where all nodes have identical degree $k$, Model \eqref{E:simp_model} can be reduced to the following model
\begin{eqnarray}{\label{E:homo_model}}
\left\{
\begin{array}{lll}
\frac{ d s}{dt}&=- \beta k s \langle \frac{k-1}{k} \rangle \left(i+\sigma a\right),\vspace{3pt}\\
\frac{ d e}{dt}&=\beta k s \langle \frac{k-1}{k}\rangle \left(i+\sigma a\right) -\epsilon e,\vspace{3pt} \\
%\frac{d a}{dt}&=\epsilon e-\nu a -\gamma^a a,  \\
%\frac{ d i}{dt}&=\nu a-\gamma i-\mu i, \\
%\frac{d r}{dt}&=\gamma i+\gamma^a a, \\
%\frac{d d}{dt}&=\mu i,
\frac{da}{dt}&=(1-\delta)\epsilon e-\gamma_a a, \vspace{3pt}   \\
\frac{ di}{dt}&=\delta\epsilon e -\left(\gamma+\mu+\xi\right) i , \vspace{3pt}\\
\frac{ dh}{dt}&=\xi i-\gamma_{h} h-\mu h, \vspace{3pt}\\
\frac{dr}{dt}&=\gamma i+\gamma_a a+\gamma_{h} h, \vspace{3pt}\\
\frac{dd}{dt}&=\mu i+\mu h,
\end{array}
\right.
\end{eqnarray}
with the initial values $s(0)=s_0$, $e(0)=e_0$, $a(0)=a_0$, $i(0)=i_0$, $h(0)=h_0$, $r(0)=0$ and $d(0)=0$.

By  Model \eqref{E:homo_model} and a direct calculation, we have
%, we can obtain that
%\begin{align*}
%%\frac{ds}{dt}+\frac{de}{dt}+\frac{da}{dt}+\frac{di}{dt}=-\gamma^a a-\left(\gamma+\mu\right)i,
%\frac{ds}{dt}+\frac{de}{dt}+\frac{da}{dt}+\frac{di}{dt}=-\gamma_a a-\left(\gamma+\mu+\xi\right)i,
%\end{align*}
%and
%\begin{align*}
%%\frac{ds}{dt}+\frac{de}{dt}+\frac{da}{dt}=-\left(\nu+\gamma^a\right)a.
%\left(1-\delta\right)\left(\frac{ds}{dt}+\frac{de}{dt}\right)+\frac{da}{dt}=-\gamma_aa.
%\end{align*}
%%Equivallently,
%%\begin{align*}
%%i&=-\frac{s'+e'+a'+i'+\gamma^a a}{\gamma+\mu},\\
%%a&=-\frac{s'+e'+a'}{\nu+\gamma_a}.
%%\end{align*}
%Equivalently,
\begin{equation} \label{eq1}
  \begin{split}
%-\left(i+\sigma a\right)=\frac{1}{\gamma+\mu}\left(\frac{ds}{dt}+\frac{de}{dt}+\frac{da}{dt}+\frac{di}{dt}\right)+\frac{\sigma-\frac{\gamma^a}{\gamma+\mu}}{\nu+\gamma^a}\left(\frac{ds}{dt}+\frac{de}{dt}+\frac{da}{dt}\right).
-\left(i+\sigma a\right)
=&\frac{1}{\gamma+\mu+\xi}\left(\frac{ds}{dt}+\frac{de}{dt}+\frac{da}{dt}+\frac{di}{dt}\right)\\
&+\left(\frac{\sigma}{\gamma_a}-\frac{1}{\gamma+\mu+\xi}\right)\left(\left(1-\delta\right)\left(\frac{ds}{dt}+\frac{de}{dt}\right)+\frac{da}{dt}\right).
  \end{split}
\end{equation}

By the first equation in Model \eqref{E:homo_model}, we further have
\begin{align}\label{eq2}
%ln\frac{s(+\infty)}{s_0}&=-\beta k \langle \frac{k-1}{k} \rangle \int_0^{+\infty}\left(i+\sigma a\right)dt.\\
%&=\beta k \langle \frac{k-1}{k} \rangle \left[\frac{s(+\infty)-y_0-i_0}{\gamma+\mu}+\frac{\sigma-\frac{\gamma^a}{\gamma+\mu}}{\nu+\gamma ^a}\left(s(+\infty)-y_0\right)\right],
&ln\frac{s(+\infty)}{s_0}
=-\beta k \langle \frac{k-1}{k} \rangle \int_0^{+\infty}\left(i+\sigma a\right)dt,%\\
%=&\beta k \langle \frac{k-1}{k} \rangle \left[\frac{s(+\infty)-y_0-a_0-i_0}{\gamma+\mu+\xi}+\left(\frac{\sigma}{\gamma_a}-\frac{1}{\gamma+\mu+\xi}\right)\left((1-\delta)\left(s(+\infty)-y_0\right)-a_0\right)\right],
\end{align}
where $s(+\infty) = \lim_{t \to \infty} s(t)$.
% denotes the fraction of susceptible individuals when $t \rightarrow \infty$ and $y_0 = s_0+e_0$.
To determine the final size of susceptible individuals, $s(+\infty)$, we set
\begin{align*}
%f(x)=s_0\exp\left\{\beta k \langle \frac{k-1}{k} \rangle \left[\frac{x-y_0-i_0}{\gamma+\mu}+\frac{\sigma-\frac{\gamma^a}{\gamma+\mu}}{\nu+\gamma ^a}\left(x-y_0\right)\right]\right\}.
f(x)
=s_0\exp&\left\{\beta k \langle \frac{k-1}{k} \rangle \left[\frac{x-y_0-a_0-i_0}{\gamma+\mu+\xi}\right.\right.\\
&+\left.\left.\left(\frac{\sigma}{\gamma_a}-\frac{1}{\gamma+\mu+\xi}\right)\left((1-\delta)\left(x-y_0\right)-a_0\right)\right]\right\},
\end{align*}
where $y_0 = s_0+e_0$.
By \eqref{eq1}, \eqref{eq2} and the definition of $f(x)$, we have
$$s(+\infty) = f(s(+\infty)).$$

%To determine the final size of susceptible individuals $s(+\infty)$, we let
%\begin{align*}
%%f(x)=s_0\exp\left\{\beta k \langle \frac{k-1}{k} \rangle \left[\frac{x-y_0-i_0}{\gamma+\mu}+\frac{\sigma-\frac{\gamma^a}{\gamma+\mu}}{\nu+\gamma ^a}\left(x-y_0\right)\right]\right\}.
%&f(x)\\
%=&s_0\exp\left\{\beta k \langle \frac{k-1}{k} \rangle \left[\frac{x-y_0-a_0-i_0}{\gamma+\mu+\xi}+\left(\frac{\sigma}{\gamma_a}-\frac{1}{\gamma+\mu+\xi}\right)\left((1-\delta)\left(x-y_0\right)-a_0\right)\right]\right\}.
%\end{align*}
It is clear that $f(x)$ is a positive, increasing, strictly convex function, and $f(s_0)<s_0$.
Thus, $f$ has a unique fixed point $s^{+}$ in the interval $(0, s_0)$, which can be calculated numerically by using the iteration method and
\begin{align*}
s^+=\lim_{m\to+\infty}{f^m(s_0)},
\end{align*}
where $f^m$ denotes composition of $f$ for $m$ times.
Then, the final size of susceptible individuals for a homogeneous network, $s(+\infty)$, can be determined by $s^{+}$.

 We now derive the final size for heterogeneous networks. Integrating the first equation in Model \eqref{E:simp_model} from $0$ to $t$, we have
\begin{align}\label{skt0}
ln\frac{s_k(t)}{s_k(0)}=-\beta k \sum_{k'}\frac{k'-1}{k'}P(k'|k)\int_0^t\left(i_{k'}+\sigma a_{k'}\right)du.
\end{align}

By summing and integrating the equations in Model \eqref{E:simp_model},
%\begin{align*}
%\frac{1}{\epsilon}\left(a_k'+\left(\nu+\gamma^a\right)a_k\right)=\frac{1}{\nu}\left(i_k'+\left(\gamma+\mu\right)i_k\right).
%\end{align*}
%Integrating it from 0 to $t$, then
%\begin{align*}
%\frac{\nu}{\epsilon}\left(a_k(t)-a_k(0)\right)+\frac{\nu(\nu+\gamma^a)}{\epsilon}\int_0^ta_kds=i_k(t)-i_k(0)+\left(\gamma+\mu\right)\int_0^ti_kds.
%\end{align*}
\begin{align}\label{eq3}
%\int_0^ti_kdu=-\frac{\nu}{(\gamma+\mu)(\nu+\gamma^a)}\left(y_k(t)-y_k(0)\right)-\frac{i_k(t)-i_k(0)}{\gamma+\mu},
\int_0^ti_kdu=-\frac{\delta}{\gamma+\mu+\xi}\left(y_k(t)-y_k(0)\right)-\frac{i_k(t)-i_k(0)}{\gamma+\mu+\xi},
\end{align}
and
\begin{align}\label{eq4}
%\int_0^t a_kdu=\frac{1}{\nu}\left(i_k(t)-i_k(0)\right)+\frac{\gamma+\mu}{\nu}\int_0^t i_k du,
\int_0^t a_kdu=-\frac{1-\delta}{\gamma_a}\left(y_k(t)-y_k(0)\right)-\frac{1}{\gamma_a}\left(a_k(t)-a_k(0)\right),
\end{align}
where $y_k(t) = s_k(t)+e_k(t)$.
We set
\begin{align*}
%f_k(t)=\frac{\nu+\sigma\left(\gamma+\mu\right)}{\left(\gamma+\mu\right)\left(\nu+\gamma^a\right)}y_k(t)+\frac{i_k}{\gamma+\mu}.
g_k(t)=\left(\frac{\delta}{\gamma+\mu+\xi}+\sigma\frac{1-\delta}{\gamma_a}\right)y_k(t)+\frac{\sigma}{\gamma_a}a_k(t)+\frac{i_k(t)}{\gamma+\mu+\xi}.
\end{align*}

By Equations \eqref{skt0},\eqref{eq3} and \eqref{eq4}, we have
\begin{align*}
&ln\frac{s_k(+\infty)}{s_k(0)}=\beta k\sum_{j=1}^n \frac{j-1}{j}P(j|k)\left(g_j(+\infty)-g_j(0)\right)\\
&= \beta k \sum_{j=1}^n\frac{j-1}{j}P(j|k)\left(\left(\frac{\delta}{\gamma+\mu+\xi}
+\sigma\frac{1-\delta}{\gamma_a}\right) s_j(+\infty)-g_j(0)\right),
\end{align*}
where
%\begin{align*}
%%f_k(0)=\frac{\nu+\sigma\left(\gamma+\mu\right)}{\left(\gamma+\mu\right)\left(\nu+\gamma^a\right)}y_k(0)+\frac{i_k(0)}{\gamma+\mu}.
%f_k(0)=\left(\frac{\delta}{\gamma+\mu+\xi}+\sigma\frac{1-\delta}{\gamma_a}\right)y_k(0)+\frac{\sigma}{\gamma_a}a_k(0)+\frac{i_k(0)}{\gamma+\mu+\xi}.
%\end{align*}
%and
$g_k(0)\ge 0, \forall k$.
Therefore, for all $k=1,...,n$, the final size of susceptible individuals satisfies
%is %can be determined by the following equations
\begin{align*}
s_k(+\infty)
=s_k(0)&\exp\left\{\beta k \sum_{j=1}^n\frac{j-1}{j}P(j|k)\cdot \right.\\
&\cdot\left.\left((\frac{\delta}{\gamma+\mu+\xi}+\sigma\frac{1-\delta}{\gamma_a})\left(s_j(+\infty)-s_j(0)\right)-w_j(0)\right)\right\},
\end{align*}
where
\begin{align*}
w_k(0)=\left(\frac{\delta}{\gamma+\mu+\xi}+\sigma\frac{1-\delta}{\gamma_a}\right)e_k(0)+\frac{\sigma}{\gamma_a}a_k(0)+\frac{i_k(0)}{\gamma+\mu+\xi}\ge 0.
\end{align*}

We define a map $G: \mathbb R^n \rightarrow \mathbb R^n$, $x= (x_j) \mapsto G(x)=\left(G_1(x),...,G_n(x)\right)^T$ by
\begin{align*}
G_i(x)
=s_i(0)&\exp\left\{\beta i\sum_{j=1}^n\frac{j-1}{j}P(j|i)\cdot \right.\\
&\cdot\left.\left((\frac{\delta}{\gamma+\mu+\xi}+\sigma\frac{1-\delta}{\gamma_a})(x_j-s_j(0))-w_j(0)\right)\right\}.
\end{align*}

To analyze the properties of $G(x)$, we shall introduce some notations.
For $Y=\left(Y_1,\dots,Y_n\right)^T$, $Z=\left(Z_1,\dots,Z_n\right)^T \in \mathbb R^n$, we denote 
 \begin{equation}\label{po}
   Y\le Z\ (\text{resp.}\  Y\ll Z)\  \text{if}\  Y_l\le Z_l \ (\text{resp.}\ Y_l < Z_l),\quad \forall l=1,\dots,n.
 \end{equation}

 Moreover, we shall claim $Y < Z$ if $Y \le Z$ and $Y \neq Z$.
 The above definition defines a partial order in $\mathbb R^n$.
 For later use, we could extend this partial order to $n\times n$ matrices as follows.
 For any $n\times n$ matrices $A, B$, we have
 $$ A \le B\ \text{if}\  Ax \le Bx,\ \forall 0 \le x\in \mathbb R^n.$$

When $0 \ll  s(0)=[s_1(0), \dots, s_n(0)]^T$ and $0 \le w(0)=[w_1(0), \dots, w_n(0)]^T$,
by the definition of $G(x)$ and partial order defined in \eqref{po}, we have
\begin{align*}
0\ll G(0)\le G(s(0)) \le s(0),
\end{align*}

Since each $G_i(x)$ is a increasing function, we have
\begin{align*}
0\ll G(0)\le \dots \le G^m(0)\le G^m(s(0)) \le \dots \le G(s(0))\le s(0),
\end{align*}
where $G^m$ is the composition function of $G$ for $m$ times.
By the monotone criterion, we obtain
\begin{align*}
0\ll \underline{s} := \lim_{m\to +\infty}G^m(0)\le \overline{s}: =\lim_{m\to +\infty}G^m(s(0)) \le s(0).
\end{align*}

Due to the continuity of $G$, $G(\underline{s})=\underline{s}$ and $G(\overline{s})=\overline{s}$.
Therefore, we have the following property \cite{wang2019further}.

\begin{lemma}
All the fixed points of $G$ in the interval $[0,s(0)]$ are contained in $[\underline{s},\overline{s}]$.
\end{lemma}

Due to the continuous differentiability of $G$,
\begin{align}\label{eq5}
\frac{\partial G_i(x)}{\partial x_j}=\beta i\frac{j-1}{j}P(j|i)\left(\frac{\delta}{\gamma+\mu+\xi}+\sigma\frac{1-\delta}{\gamma_a}\right)
G_i(x)
%\dot= b_{ij}G_i(x)
\end{align}
for any $x\in \mathbb R^n$ and $1\le i,j \le n$.
Moreover, we shall simply write \eqref{eq5} in terms of the matrix form by
\begin{align*}
DG(x):= \frac{\partial G(x)}{\partial x}={\rm diag}\left(G_1(x),\cdots, G_n(x)\right)B,
\end{align*}
where $B=[b_{ij}]$ and $b_{ij}=\beta i
\frac{j-1}{j}\left(\frac{\delta}{\gamma+\mu+\xi}+\sigma\frac{1-\delta}{\gamma_a}\right)P(j|i)$.

By the monotony of $G$, $DG$ is also monotonous, i.e., $DG(x)\le DG(y)$ for any $0\le x\le y\le s(0)$.
By utilizing the properties of $w(x)$ and $G(x)$, we can obtain the following theorem.
\begin{theorem}\label{h_network_root}
Assume that the network is connected, we have
\begin{itemize}
\item[\rm{(1)}] $w(0)=0$ if and only if $G(s(0))=s(0)$;
\item[\rm{(2)}] when $w(0)>0$, $G$ has a unique fixed point $s^{++}$ satisfying $0\ll s^{++}<s(0)$.
\end{itemize}
%As denoted above, $s(t)=[s_1(t), \dots, s_n(t)]'$.
\end{theorem}

The proof of Theorem \ref{h_network_root}  directly follows the proof in \cite{wang2019further}. Hence, the final size of susceptible individuals for a heterogeneous network, $s(+\infty)$, can be determined by $s^{++}$ to quantify the number of susceptible individuals left theoretically.

%The early epidemic threshold is determined by the basic reproduction number, $R_0$. This dimensionless number is defined as the average number of secondary cases produced by an infected individual in a completely susceptible population.19
%The basic reproduction number can be used to determine whether an epidemic will spread or die out.  We derived an explicit expression for R0 using the next-generation matrix approach20,
%$ R_0=\beta  \Big(\frac{1-\delta}{\gamma_a}\sigma+\frac{\delta}{\gamma +\mu+\xi}\Big) (<k^2>-<k>) .$
%Here, $\beta\sigma(1-\delta)/\gamma_a$ and $\beta\delta(\gamma+\mu+\xi$
%represent the average numbers of secondary cases produced by an asymptomatically infected individual and a symptomatically infected individual in a homogeneously mixed population, respectively. The term $(<k^2>-<k>)/<k>$ represents the average excess degree of nodes in the network.21
%When the heterogeneity among nodes increases, then the average number of cases produced by an infectious individual (the reproductive number) increases and the epidemic is more likely to spread.

\section{Parameter estimation and model based forecasting}

We parameterized the model with reported data on COVID-19 cases and presented forecasts of the epidemic trends for the three areas.

\subsection{Fitting reported confirmed cases}

We simulated the spread of COVID-19 in Wuhan, Toronto, and Italy on the Watts-Strogatz network with degree $k_{min}= 1$ and $k_{max}= 10$.

The study period for Wuhan starts from January 11, 2020, after the confirmed cases were reported, the public becomes aware of the infection and most people are trying to avoid gathering. The study period starts from January 26 for Toronto  and from January 31 for Italy. In Toronto and Italy, usually people do not gather, especially after lockdown on Wuhan city, the awareness of avoiding exposure to the virus is increasing. Most people stay home during the study period, and the family sizes  in Wuhan, Toronto, and Italy on average are all  around $3$. Therefore, the range of the node degrees is assumed to be between $1$ and  $10$.

The Watts-Strogatz model starts with a ring of $N$ vertices in which each vertex is  connected to its $2m$ nearest neighbors ($m$ vertices clockwise and $m$ counterclockwise). Each edge is connected to a clockwise neighbor with probability $p$ and preserved with probability $1-p$ \cite{Barrat2008}, where the degree distribution is $$ P(k)=\sum_{n=0}^{\min(k-m, m)} \begin{pmatrix}
  m\\
  n
\end{pmatrix}(1-p)^n p^{m-n}\frac{(pm)^{k-m-n}}{(k-m-n)!} e^{-pm}.$$
When $p  \rightarrow 1 $, the expression reduces to a Poisson distribution as follows $$ P(k)=\frac{m^{k-m}}{(k-m)!}e^{-m}.  $$
In the simulations, we used this degree distribution.

The total number of nodes  for Wuhan, Toronto, and Italy are  $11081000$, $5928000$, and $59430000$ as shown in Table \ref{Parametes_Wuhan}, Table \ref{Parametes_Toronto}, and Table \ref{Parametes_Italy}, respectively. We parameterized the model using the MCMC approach \cite{haario2006dram} by MATLAB R2016a according to the number of newly confirmed cases and the cumulative number of cases reported by the Health Commission of Hubei Province \cite{Health_Commission2020} and WHO \cite{WHO_report2020}.

The rate at which the fraction of the cumulative number of cases changes is $dc_k/dt=\xi i_k$, where $c_k(t)$ represents the fraction of the cumulative number of infected individuals with degree $k$. The number of newly infected can be expressed as $$P_k=[c_k(t)-c_{k-1}(t)]N_k,$$ where $P_k$ represents the number of new cases with degree $k$, and $N_k$ represents the total number of individuals with degree $k$. We run the MCMC simulation for $20000$ iterations to fit the value of $P_k$.

Zhou et al. showed that the median time from illness onset (i.e., before admission) to discharge was 22 days (IQR 18-25), whereas the median time to death was 18.5 days with IQR between $15$ and $22$ days \cite{zhou2020clinical}. We assume an exponential distribution for the time to recovery for asymptomatically infected individuals, symptomatically  infected individuals, and hospitalized individuals.
This results in the recovery rates $\gamma_a=\gamma=\gamma_h=1/22$ per day, and the mortality rate, $\mu$ is $1/18.5$ per day.
The incubation period of COVID-19 is around $7$ days \cite{kucharski2020early},
resulting in the progression rate $\epsilon=1/7$.
Qiu et al. reported that around $30\%-60\%$ of people infected with COVID-19 are asymptomatic or only have mild symptoms, and their transmissibility is lower, but still significant \cite{qiu2020covert}.
Thus, we assume that the probability that an infected individual is asymptomatic is $1-\delta=0.6$, and $\sigma=1$ for simulations.

We divided the Wuhan epidemic into four phases according to the reported data \cite{wang2020phase}.
 The first phase is before lockdown on Jan 23, 2020. The second phase is between Jan 24, 2020 and Feb 1, 2020 when the hospitals were short of beds. The third phase is between Feb 2, 2020 and Feb 6, 2020 when the Thunder God Mountain Hospital (TGMH) and Fire God Mountain Hospital (FGMH) were put into use. The fourth phase began when door-to-door screening was implemented on Feb 7, 2020 and TGMH, FGMH, and Mobile Cabin Hospitals (MCH) were put into use.

The study period for Toronto (Canada) was decomposed into two phases, namely, the period before Mar 18 and the period after Mar 18 when the city announced the emergence and schools and universities in Toronto were closed on Mar 18.

The study period for Italy was divided into two phases. The early epidemic phase was between Jan 31, 2020 and Mar 8, 2020 when the infection was spreading through the northern provinces. The second period begins on Mar 9, 2020 when the national lockdown started.

\subsection{Predicting future epidemic trends}
%\subsection{The impact of mitigation strategies}

%Here are two sample references: \cite{Feynman1963118,Dirac1953888}.

%\section{Results}\label{Results}
%We forecasted the epidemic trends and estimated the impact of various mitigation strategies.
%\subsection{The infection dynamics}
%\subsubsection{The infection dynamics in Wuhan}
The parameters and initial conditions of simulations for Wuhan on the WS network are shown in Table \ref{Parametes_Wuhan}. The probability of transmission through adequate contact is estimated by MCMC. The $5000$ realizations of the basic reproduction numbers derived for Wuhan using the parameter values listed in Table \ref{Parametes_Wuhan} are shown in Table \ref{Reproducton_numbers}.

\begin{figure}[http]
\caption{Fitting the number of reported new cases and the cumulative number of reported cases between Jan 11, 2020 and Mar 31, 2020 for Wuhan on Watts-Strogatz network.(A) Fitting the number of reported new cases on the Watts-Strogatz network. (B) Fitting the cumulative number of reported cases on the Watts-Strogatz network.
}\label{Wuhan}
\centering
\includegraphics{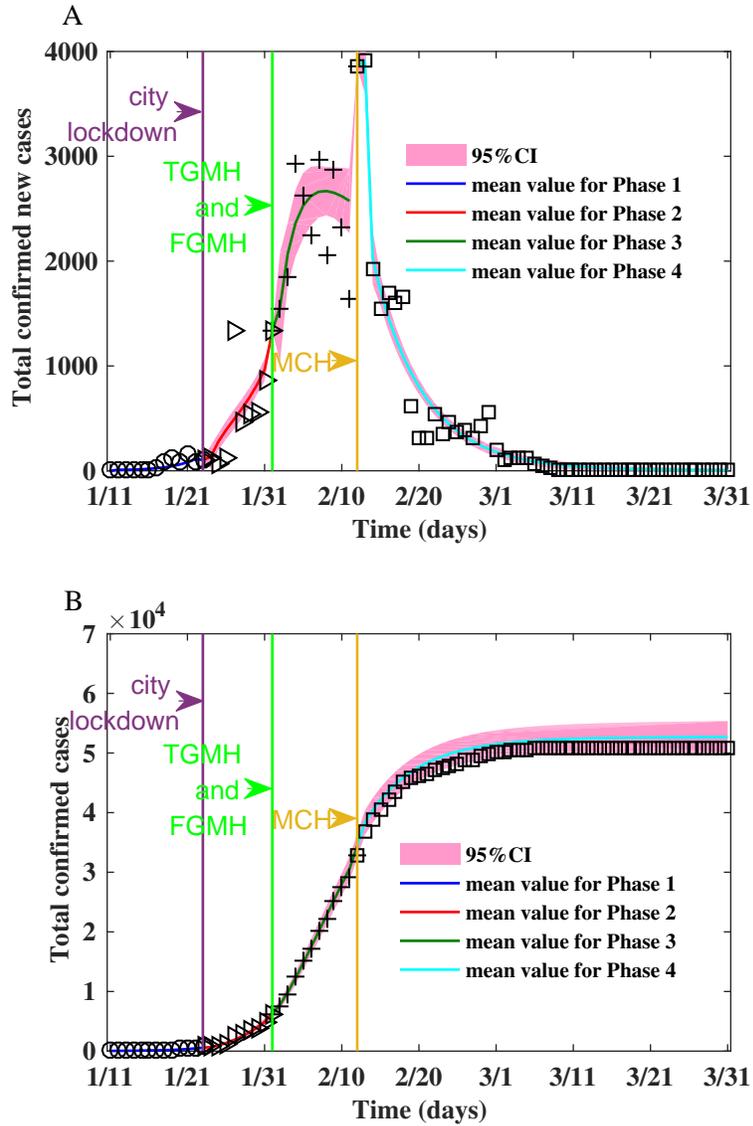}
\end{figure}

From Jan 11 to Mar 31, we estimate that the mean reproduction number on the WS network decreases from $3.41$ in the first phase to $5.34\times10^{-3}$ in the fourth phase. The epidemic on the WS network is shown in Figure \ref{Wuhan}. Up to Jan 23, 2020 when Wuhan lockdown started, the estimated epidemic size is $3.96\times10^6$. During the second stage, after the lockdown of Wuhan and before the TGMH and FGMH were put into use, the predicted final size is $2.17\times10^6$. Thus, the lockdown of Wuhan reduced the expected final size by $45.22\%$. During the third stage, after TGMH and FMGH were put into use, the final size is $1.02\times10^5$. Hence, the city lockdown and the usage of  TGMH and FGMH reduced the final size by $97.42\%$. During the fourth stage, after MCH was put into use, the predicted final size is $51269$, and the expected final size of infection is reduced by $98.70\%$ due to the increase of healthcare capacity.

The variability of the numbers of confirmed new cases is consistent with the variability of the reproduction numbers listed in Table \ref{Reproducton_numbers}. In the first two phases, the epidemic spread rapidly with larger reproduction numbers that are larger than 1, and the numbers of infected cases increase. In the last two phases, the spread of disease is controlled, and the reproduction numbers are smaller than 1. In the third phase, because a large number of cases are confirmed by door-to-door screening and expanded  healthcare capacity, the cumulative number of confirmed cases increased. On the other hand, the epidemic will die out because the reproduction number is less than one. In the fourth phase, the spread of the disease has been under control with the reproduction number being less than one. Hence, the number of new cases decreases.

\begin{figure}[http]
\centering
\caption{Fitting the number of reported new cases and the cumulative number of reported cases for Toronto on the Watts-Strogatz network. (A) Fitting the number of reported new cases on the Watts-Strogatz network. (B) Fitting the cumulative number of reported cases on the Watts-Strogatz network.}
\label{Toronto}
\includegraphics{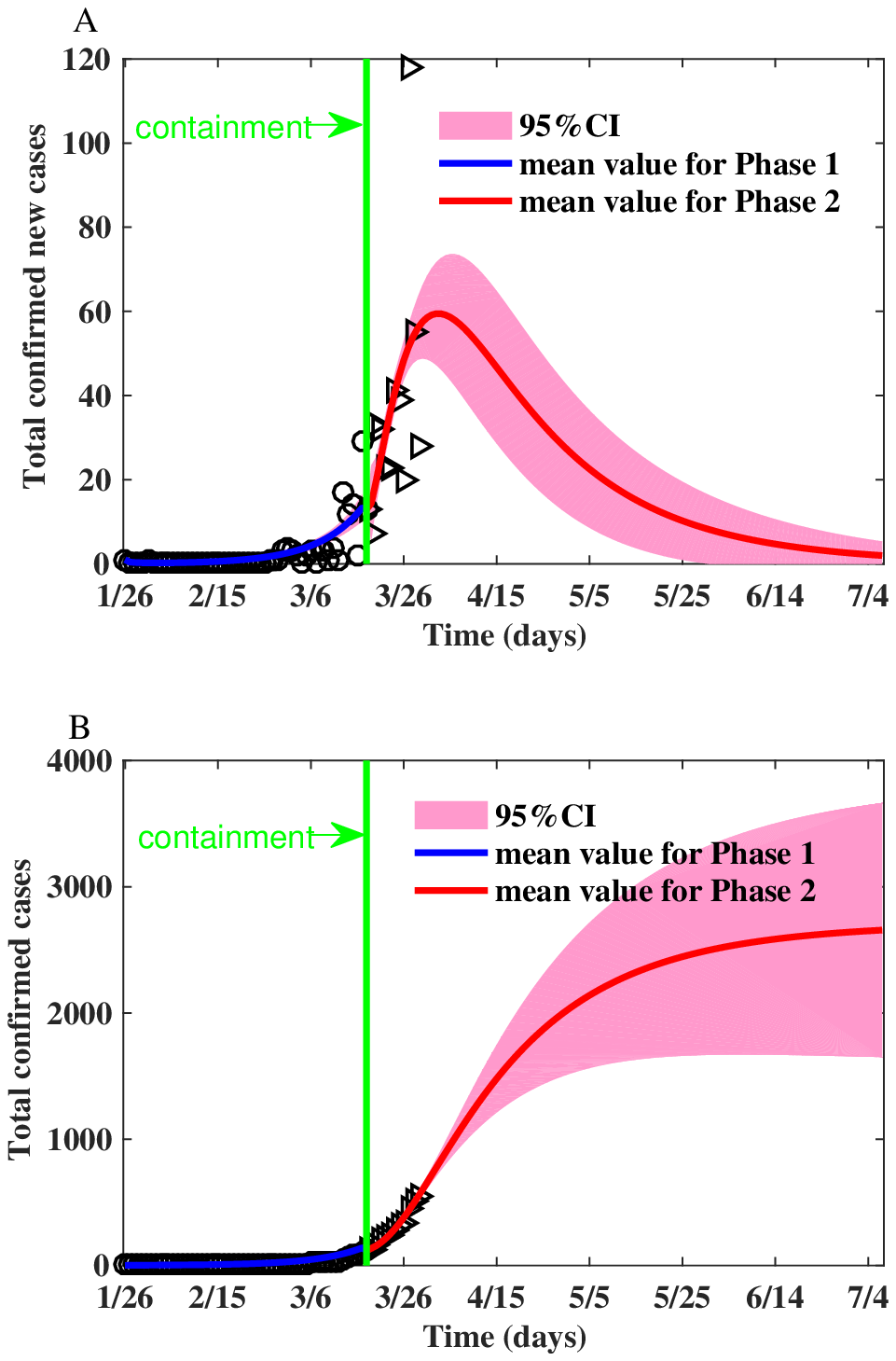}
\end{figure}

%\subsubsection{The infection dynamics in Toronto}
The parameters and initial conditions of simulations for the GTA are shown in Table \ref{Parametes_Toronto}.  The $5000$ realizations of the basic reproduction numbers derived for Toronto using the parameter values listed in Table \ref{Parametes_Toronto} are shown in Table \ref{Reproducton_numbers}. The reproduction numbers are much smaller due to social distancing policy, school closure, as well as behavior changes. The summary of the simulations is shown in Table \ref{Peak_Toronto} and Table \ref{VaringPeak_Toronto}. Simulation results are shown in Figure \ref{Toronto}. The peak size is 60.19 (95\%CI: 47.42-72.97), the peak time is Apr 2 (95\%CI: Mar 29-Apr 7), and the final size is $2712$ (95\%CI: 1603-3820).

\begin{figure}[http]
\centering
\caption{ Fitting the number of reported new cases and the cumulative number of reported cases for Italy on the Watts-Strogatz network. (A) Fitting the number of reported new cases on the Watts-Strogatz network. (B)  Fitting the cumulative number of reported cases on the Watts-Strogatz network.
}
\label{Italy}
\includegraphics{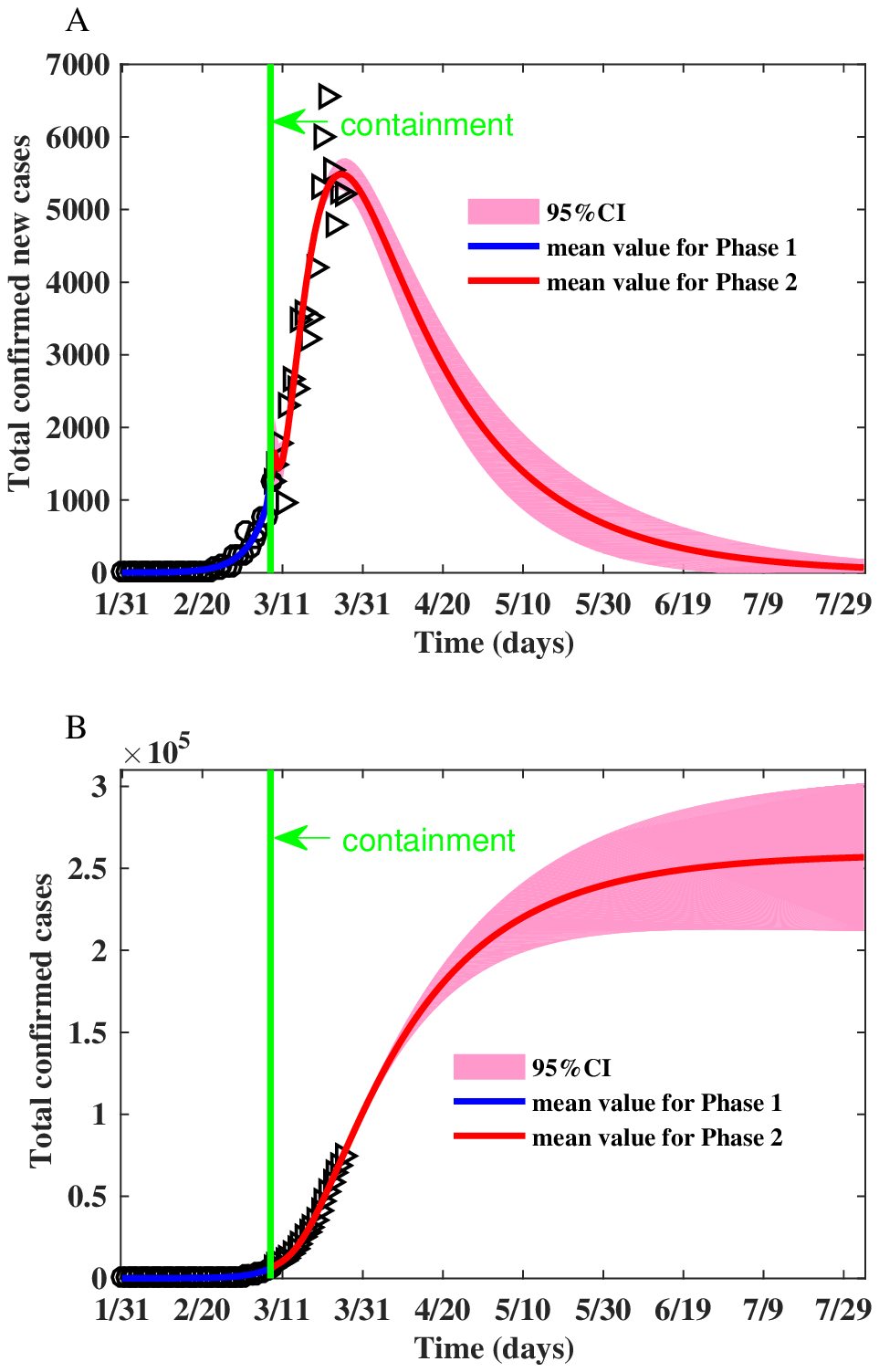}
\end{figure}

%\subsubsection{The infection dynamics in Italy}
The parameters and initial condition of simulations for Italy is shown in Table \ref{Parametes_Italy}. The $5000$ realizations of the basic reproduction numbers derived for Italy using the parameter values listed in Table \ref{Parametes_Italy} are shown in Table \ref{Reproducton_numbers}. The reproduction numbers in the second phase are much smaller than that in the first phase due to the awareness of the severity of the epidemic. The summary of the simulation results is shown in Table \ref{Peak_Italy} and Table \ref{VaringPeak_Italy}. Figure \ref{Italy} shows that the peak number of new cases is 5492 (95\%CI: 5277-5708) on Mar 26 (95\%CI: Mar 24-Mar 27), and the final size is $2.59\times 10^5$ (95\%CI: $2.10\times10^5-3.08\times10^5$).

\section{The impact of mitigation strategies}
%\subsubsection{The impact of mitigation strategies on Wuhan}

  The close contacts identified by contact tracing will be quarantined due to exposure to COVID-19 to see if they become sick.
To evaluate the impact of mitigation strategies on the spread of COVID-19, Model \eqref{E:mainmodel} is rewritten as follows

\begin{eqnarray}{\label{E:mainmodel1}}
\left\{
\begin{array}{lll}
\frac{ ds_k}{dt}&=- \beta k s_k\sum_{k'}[1-(1-\frac{k'-1}{k'}P(k'|k)i_{k'})(1-\frac{k'-1}{k'}P(k'|k)\sigma a_{k'})]-qs_k+\lambda{sq}_k,\\ \vspace{3pt}
\frac{ d{sq}_k}{dt}&=qs_k-\lambda{sq}_k, \\\vspace{3pt}
\frac{ de_k}{dt}&=\beta k s_k\sum_{k'}[1-(1-\frac{k'-1}{k'}P(k'|k)i_{k'})(1-\frac{k'-1}{k'}P(k'|k)\sigma a_{k'})] -\epsilon e_k,  \\
\frac{da_k}{dt}&=(1-\delta)\epsilon e _k-\gamma_a a_k,    \\ \vspace{3pt}
\frac{ di_k}{dt}&=\delta\epsilon e _k -\gamma i_k-\mu i_k-\xi i_k , \\ \vspace{3pt}
\frac{ dh_k}{dt}&=\xi i_k-\gamma_{h} h_k-\mu h_k, \\ \vspace{3pt}
\frac{dr_k}{dt}&=\gamma i_k+\gamma_a a_k+\gamma_{h} h_k, \\ \vspace{3pt}
\frac{dd_k}{dt}&=\mu i_k+\mu h_k,
\end{array}
\right.
\end{eqnarray}
where $sq_k=SQ_k/N_k$ represents the fraction of quarantined individuals with degree $k$. The  parameter $q$ represents the rate at which susceptible individuals are quarantined, and $\lambda$ represents the rate at which the quarantined and uninfected close contacts transfer to the susceptible compartment again. In the simulations, we let $\lambda=1/14$ to approximate a mean time of 14 days in the exposed state.

\begin{figure}[http]
\centering
\caption{The impact of the variability on the healthcare capacity on the spread of the epidemic in Wuhan on the Watts-Strogatz network.
}
\label{health_Wuhan}
\includegraphics{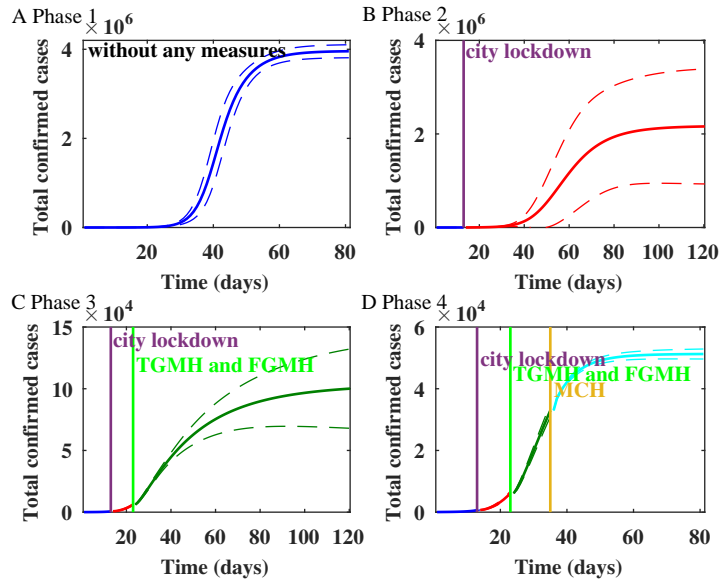}
\end{figure}

For Wuhan, the cumulative number of infected individuals after lockdown and TGMH, FGMH, as well as MCH were put into use are shown in Figure \ref{health_Wuhan}. The results show that the lockdown and the increase in healthcare capacity are effective in controlling the numbers of confirmed cases.

\begin{figure}[http]
\centering
\caption{The impact of mitigation strategies on the spread of COVID-19 epidemic in Toronto on the Watts-Strogatz network. In this figure and the following figure, the dashed lines represent $95\%$ confidence intervals. In (A) and (B), the red, purple, and green lines represent that the transmission rates are unchanged, reduced by $20\%$, and reduced by $40\%$, respectively. In (C) and (D), the red, purple, and green lines represent the rate of quarantine, $q=0$, 1/8, and 1/4, respectively. In (E) and (F), the red, purple, and green, and light blue lines represent that the node degrees of symptomatically infected individuals are reduced by 0, 1, 2, and 3, respectively.  (A) The number of newly infected individuals after reducing the transmission rates by personal protection and social distancing. (B) The cumulative number of infected individuals after reducing the transmission rates by personal protection and social distancing. (C) The number of newly infected individuals after close contacts are quarantined. (D) The cumulative number of infected individuals after close contacts are quarantined. (E) The number of newly infected individuals after the node degrees of symptomatically infected individuals are reduced. (F) The cumulative number of infected individuals after the node degrees of symptomatically infected individuals are reduced.
}
\label{health_Toronto}
\includegraphics{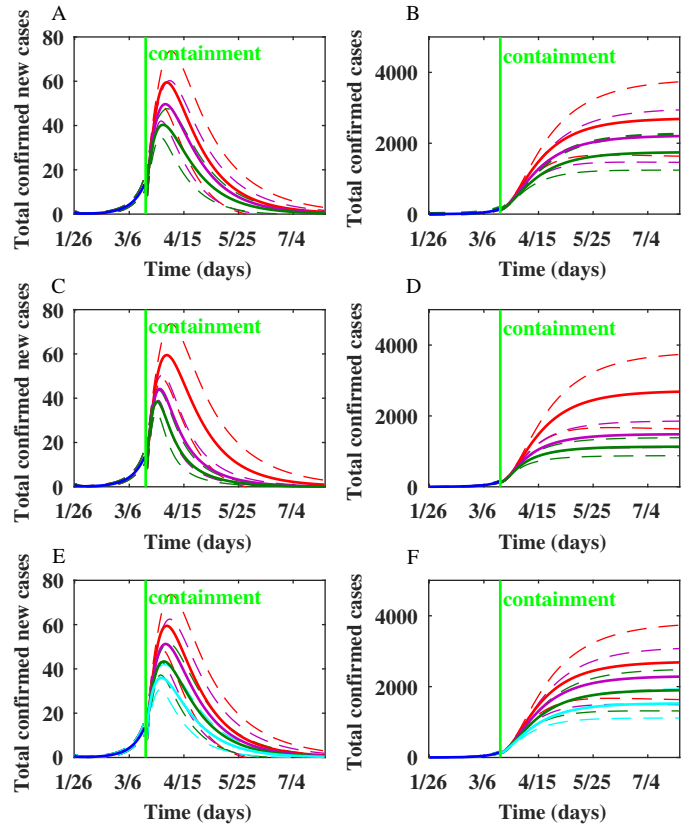}
\end{figure}

%\subsubsection{The impact of mitigation strategies on Toronto}
For Toronto, the number of newly infected individuals and the cumulative number of infected individuals produced on the WS network after implementing additional containment strategies besides school closure are shown in Figure \ref{health_Toronto}. We simulated the scenarios of implementing various containment strategies for Toronto. Simulation results showed that personal protection, reducing the node degrees of symptomatically infected individuals, and quarantine of close contacts are effective in reducing the peak epidemic size and final epidemic size. Reducing the transmission rate $\beta$,  by $x\%$ also reduces $R_0$ by $x\%$. When  $\beta$  is reduced by $20\%$ by personal protection or social distancing, the peak occurs one day earlier, and the final epidemic size is reduced by around $18\%$. When $\beta$  is reduced by $40\%$, the peak occurs two days earlier, and the final epidemic size is reduced by around $33.3\%$. When $q=1/8$, the peak occurs four days earlier, and the final epidemic size is reduced by $45.21\%$. When $q=1/4$, the peak appears five days earlier, and the final epidemic size is reduced by $58.22\%$. When the node degrees of symptomatically infected individuals are reduced by $1$, $2$, and $3$, the number of new cases produced per day at the peak  is reduced by $13.74\%$, $26.93\%$, and $39.18\%$. The final epidemic size is reduced by $15.15\%$, $29.65\%$, and $43.55\%$ when the node degrees of symptomatically infected individuals are reduced by $1$, $2$, and $3$, respectively.

\begin{figure}[http]
\centering
\caption{The impact of mitigation strategies on the spread of COVID-19 epidemic in Italy on the Watts-Strogatz network. (A) The number of newly infected individuals after reducing the transmission rates by personal protection and social distancing. (B) The cumulative number of infected individuals after reducing the transmission rates by personal protection and social distancing. (C) The number of newly infected individuals after close contacts are quarantined. (D) The cumulative number of infected individuals after close contacts are quarantined. (E) The number of newly infected individuals after the node degrees of symptomatically infected individuals are reduced. (F) The cumulative number of infected individuals after the node degrees of symptomatically infected individuals are reduced.
}
\label{health_Italy}
\includegraphics{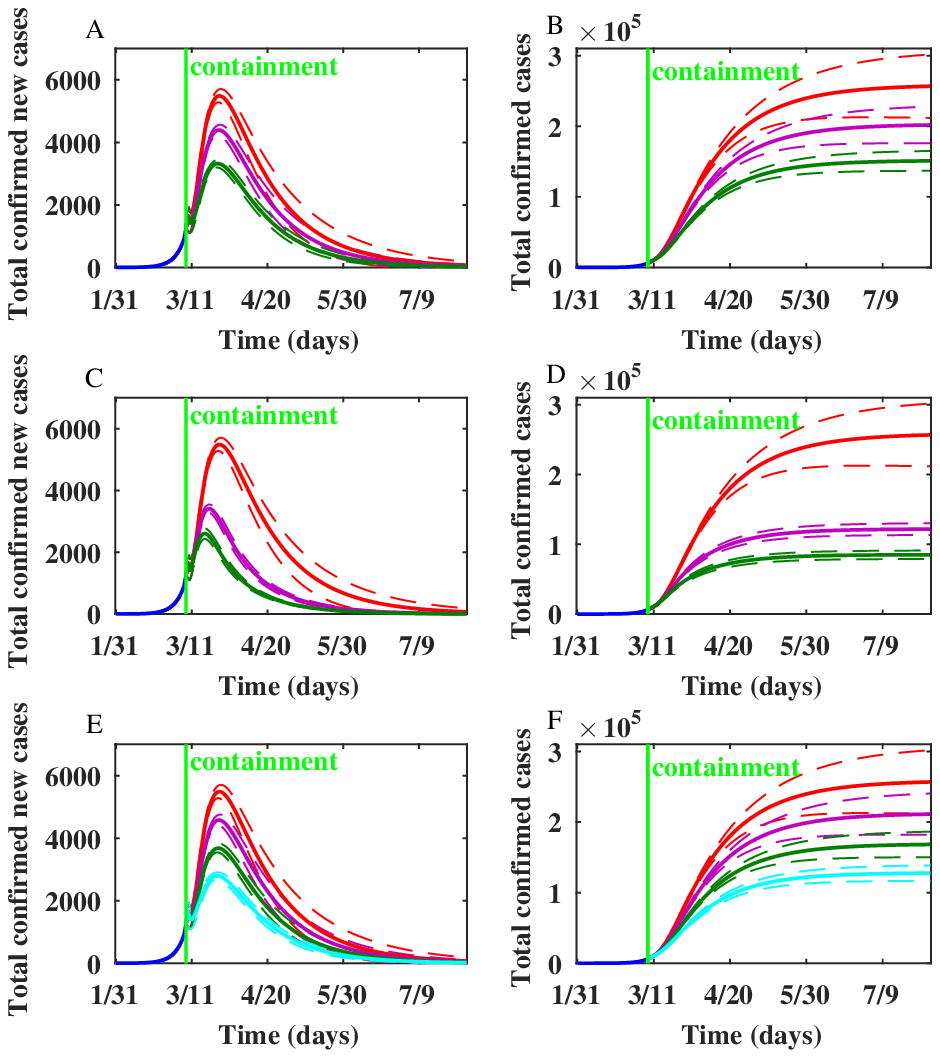}
\end{figure}

%\subsubsection{The impact of mitigation strategies on Italy}
For Italy, the number of newly infected individuals and the cumulative number of infected individuals simulated on the WS network after implementing hypothetical containment strategies are shown in Figure \ref{health_Italy}.
Various scenarios of implementing mitigation strategies showed that the peak epidemic size and final epidemic size in Italy are greatly reduced by personal protection, social distancing, behavior change of symptomatically infected individuals, and quarantine. The simulations show that the peak would have arrived earlier if the containment had been intensified.

When the probability of contact transmission coefficient $\beta$, is reduced by $20\%$ by personal protection or social distancing, the peak occurs one day earlier, and the final epidemic size is reduced by $21.56\%$. When $\beta$ is reduced by $41.44\%$, the peak occurs one day earlier, and the final epidemic size is reduced by around $40\%$.

When $q=1/8$, the peak occurs six days earlier, and the final epidemic size is reduced by $52.87\%$. Yet, when $q=1/4$, the peak occurs eight days earlier, and the final epidemic size is reduced by $67.12\%$. When the node degree of symptomatically infected individuals is reduced by $1$, $2$, and $3$, the number of new cases produced per day at the peak is reduced by $16.50\%$, $32.93\%$, and $49.11\%$, respectively. The final epidemic size is reduced by $17.90\%$, $34.70\%$, and $50.51\%$ when the node degrees of symptomatically infected individuals are reduced by $1$, $2$, and $3$, respectively.

\section{Summary and Discussions}\label{Discussion}
Modelling the dynamics of COVID-19 epidemics and assessment of mitigation strategies could be instrumental to public health agencies for surveillance and healthcare planning. For the models to be reliable, the simulated epidemic must account for the stochastic and heterogeneous contact among individuals. Hence, we developed a network model that captured the contact heterogeneity among individuals. We applied the model to analyze the transmission potential, and mitigation strategies for curbing the spread of COVID-19 epidemics in the cities of Wuhan, China and Toronto, Canada, and in the Italian Republic. The epidemic threshold derived from our network model  can be used to predict the risks of spreading scenarios.  We also provided an explicit expression of the final epidemic size, which facilitates estimating the scale of an outbreak for any region of interest.  Our results provide insights in defining a mathematical framework for the analysis and containment of epidemic transmission in the real world.

The flexible network model framework can simulate a wide range of mitigation strategies can be examined by the flexible model framework. It can be extended to quantify the effectiveness of personal protection, social distancing, reducing the node degree of infected individuals, and quarantine on the dynamics of epidemics in different regions. When the mitigation strategy is intensified, the model predicts that the number of new cases peaks earlier and the final epidemic size is greatly reduced. 

%We assume that the range of the node degree is between one and ten for each network in the absence of real contact tracing data. In reality, the range of the degree may vary for different regions.

%In particular, there are reported cases of super spreaders where a single individual has infected over $40$ other people.  This could be captured by simply increasing the upper bound for the distribution of the node degrees.

The social contact network structure and parameter values  determine the
transmission and epidemic course of such an emerging infectious disease. We choose the Watts-Strogatz to approximate  real social networks, when the exact contact tracing data is unavailable. We assumed that the range of the node degree is between one and ten for each network in the absence of real contact tracing data, that is, on average each day an infected person would have between one and ten contacts where they could transmit the infection to another person.  In the real world, the range of the degree will depend on the distribution of the household sizes of the region and time being studied. Moreover, the network structure can be altered by behavior change of individuals during epidemics. When this happens, the network structure can be adapted in our model to  predict the impact of these changes on the epidemic threshold, epidemic peak value, peak time, stopping time, and final size of infected population.

The epidemics for the three places under study were fitted very well by our model with a small confidence interval. Hence, the forecasts by the model can be reliable. We did not provide the stopping time since too many uncertainties may affect the duration of the epidemics. As shown in the simulations, the transmission dynamics for four phases in Wuhan are quite different due to the variability on the intensity of interventions, the availability of healthcare facilities, as well as the utilization of  personal protective equipment (PPE). The dynamics in the first phase is quite different from that in  the second phase for Toronto. The same phenomenon is observed in Italy.

At the early stage, almost no interventions were implemented, and the public was not aware of or did not pay much attention to the severity of the highly contagious disease. With the increase of the number of reported confirmed cases and with the aid of social media, the public becomes aware of the severe consequence and has increased the level of personal protection and have avoided gathering, so that the reproduction number decreases and the estimated epidemic size declines by reducing the node degree of the network. Similarly, after applying the mitigation measures in Italy on March 8 and closing all schools in Toronto on March 18, the epidemics tend to be under control.

Hence, social distancing, self-isolation, quarantine, the utilization of PPE, and other measures of avoiding exposure to the virus can greatly reduce the size of infection during the COVID-19 outbreak. Therefore, it is essential to raise the awareness of these countermeasures % personal protection, social distancing, self-isolation, as well as other possible protective measures
to avoid contact between individuals. The possibility of recurrent outbreaks of the disease cannot be overstated. Even if the number of new cases is declining, it is still necessary to continue taking protective measures to prevent the occurrence of future outbreaks. The social media should warn the public not to relax their vigilance against the contagion of such a highly infectious disease.

\begin{table}
\caption{\label{Parametes_Wuhan} \textbf{The parameter values and initial condition of four-phase simulations for Wuhan.
}Phase 1 is between Jan 11 and Jan 23, Phase 2 is between Jan 23 and Feb 1, Phase 3 is between Feb 1 and Feb 12, and Phase 4 is between Feb 12 and Mar 31. }
\centering
%\begin{tabular}{p{2.65cm}|p{2cm}|p{2cm}|p{3.6cm}|p{1.59cm}}
\begin{tabular}{p{2.65cm}|p{2cm}|p{2cm}|p{3.6cm}|p{1.59cm}}
\hline
Parameter& Mean value & Std& 95\% CI &References \\
\hline
$\beta$ (Phase 1)& $0.04644$ & $2.83\times10^{-3}$ &[$0.0409$, $0.0520$] & MCMC\\
$\beta$ (Phase 2)& $0.01597$ & $4.79\times10^{-3}$ &[$6.59\times10^{-3}$, $0.0254$] & MCMC\\
$\beta$ (Phase 3)& $2.90\times 10^{-4}$ & $2.24\times10^{-4}$ &[$0$, $7.29\times10^{-4}$] & MCMC\\
$\beta$ (Phase 4)& $7.30\times 10^{-5}$ & $ 7.27\times10^{-5}$ &[$0$, $2.16\times10^{-4}$] & MCMC\\
$\xi$ (Phase 1)& $0.8550$ & $0.1220$ &[$0.6159$, $1.0942$] & MCMC\\
$\xi$ (Phase 2)& $0.1499$ & $0.1557$ &[$0$, $0.4552$] & MCMC\\
$\xi$ (Phase 3)& $0.3369$ & $0.1089$ &[$0.1234$, $0.5504$] & MCMC\\
$\xi$ (Phase 4)& $0.8764$ & $0.0836$ &[$0.7124$, $1.0403$] & MCMC\\
$\sum_k N_{k}(0)$& $11081000 $ & $-$ &$-$ & \cite{HPBA2020}\\
$\sum_k S_{k}(0)$& $11080770$ & $-$ &$-$ &Calculated\\
$\sum_k E_{k}(0)$& $200.26$ & $18.88$ &[$163.25$, $237.26$] & MCMC\\
$\sum_k A_{k}(0)$& $17.81$ &$4.13$ & [$9.71$, $25.90$]& MCMC\\
$\sum_k I_{k}(0)$& $11.78$ & $ 6.06$ &[$0$, $23.66$] &  MCMC\\
$\sum_k H_{k}(0)$& $41$ &$-$ &$-$& \cite{Health_Commission2020} \\
$\sum_k R_{k}(0)$& $0$ & $-$ &$-$&  Estimated\\
$\sum_k D_{k}(0)$ &  $0$ & $-$ &$-$ &Estimated \\
\hline
\end{tabular}
\end{table}

\begin{table}
\caption{\label{Reproducton_numbers} \textbf{Basic reproduction numbers computed by MCMC on the WS network. }}
\centering
\begin{tabular}{p{1.6cm}|p{2.8cm}|p{2cm}|p{2cm}|p{2.8cm}}
\hline
Location&Period	& Mean value & Standard derivation & 95\% CI\\
\hline
 \multirow{4}{*} {Wuhan}&Jan 11-Jan 23&	3.4074&	0.2099&	[2.9959, 3.8188]\\
 \cline{2-5}
& Jan 23-Feb 1 &	1.3065&	0.3976&	[0.5273, 2.0858]\\
 \cline{2-5}
&Feb 1-Feb 12 &	0.0221&	0.0170&	[0, 0.0555]\\
 \cline{2-5}
&Feb 12-Mar 31&	$5.35\times10^{-3}$&	$5.34\times10^{-3}$&	[0, 0.0158]\\
\cline{2-5}
\hline
 \multirow{2}{*}{Toronto}& Jan 26-Mar 18&	0.6416&	0.0867&	[0.4716, 0.8116]\\
   \cline{2-5}
&Mar 18-Mar 29 &	0.0115&	0.0151&	[0, 0.0412]\\
\hline
 \multirow{2}{*} {Italy}&Jan 31-Mar 8&	1.4763&	0.0984&	[1.2834, 1.6691]\\
  \cline{2-5}
&Mar 8- Mar 26& 0.0359&	0.0185&	[0, 0.0721]\\
\cline{2-5}
\hline
\end{tabular}
\end{table}

\begin{table}
\caption{\label{Parametes_Toronto} \textbf{The parameter values and initial condition of simulations for Toronto. }Phase 1 is from Jan 26 to Mar 18, and Phase 2 is from Mar 18  to Mar 29.}
\centering
\begin{tabular}{p{2.2cm}|p{2cm}|p{2cm}|p{3.2cm}|p{1.59cm}}
\hline
Parameter& Mean value & Std& $95\%$ CI &References \\
\hline
$\beta$ (Phase 1) &	$7.95\times 10^{-3}$	&$1.12\times 10^{-3}$&	$[5.76\times 10^{-3}, 0.01]$ &MCMC\\
$\beta$ (Phase 2) & $1.38\times 10^{-4}$ & $1.79\times 10^{-4}$&$[0, 4.90\times 10^{-4}]$&MCMC\\
$\xi$ (Phase 1) &	0.1421&	0.0734&	[0, 0.2858]&	MCMC\\
$\xi$ (Phase 2) &	0.1140&	0.0709	&[0, 0.2530]&	MCMC\\
$\sum_k N_k(0)$&5928000 &-&-& \cite{Torontopopulation}\\
$\sum_k S_k(0)$&5927990&-&-&Calculated\\
$\sum_k E_k(0)$&4.12&4.83&[0, 13.58]&MCMC\\
$\sum_k A_k(0)$&2.43	&2.22&[0, 6.78]&MCMC\\
$\sum_k I_k(0)$&3.22	&2.89&[0, 8.88]	&MCMC\\
$\sum_k H_k(0)$&1&	-	&-&	\cite{WHO_report2020}\\
$\sum_k R_k(0)$&0&	-	&-	&Estimated\\
$\sum_k D_k(0)$&0&	-	&-	&Estimated\\
\hline
\end{tabular}
\end{table}

\begin{table}
\caption{\label{Peak_Toronto} \textbf{The peak number of new cases, peak time, and final epidemic size after containment strategies are implemented in Toronto.}}
\centering
\begin{tabular}{p{1.5cm}|p{3.5cm}|p{4.4cm}|p{3.3cm}}
\hline
Scenarios&Peak size (95\%CI)&	Peak time (95\%CI)	&Final size (95\%CI)\\
\hline
$\beta$   & 60.19 (47.42, 72.97) & Apr 2 (Mar 29, Apr 7) & 2712 (1603, 3820)\\
$0.8\beta$&	50.30 (41.03, 59.57)&Apr 1 (Mar 28, Apr 6)&2217 (1451, 2984)\\
$0.6\beta$&	40.94 (34.55, 47.32)&Mar 31 (Mar 26, Apr 4)&1751 (1239, 2262)\\
$q=0$	&   60.19 (47.42, 72.97)&Apr 2 (Mar 29, Apr 7)	&2712 (1603, 3820)\\
$q=1/8$&	44.66 (39.12, 50.21)&Mar 28 (Mar 26, Mar 31)	&1486 (1111, 1861)\\
$q=1/4$&	38.79 (33.86, 43.72)&Mar 27 (Mar 25, Mar 29)	&1133 (878, 1388)\\
\hline
\end{tabular}
\end{table}

\begin{table}
\caption{\label{VaringPeak_Toronto} \textbf{The peak number of new cases, peak time, and final epidemic size for Toronto when varying node degrees of symptomatically infected individuals.}}
\centering
\begin{tabular}{p{1.1cm}|p{3.5cm}|p{4.2cm}|p{3.5cm}}
\hline
Degree &Peak size (95\%CI)&	Peak time (95\%CI)	&Final size (95\%CI)\\
\hline
$k$	&       60.19 (47.42, 72.97)&	Apr 2 (Mar 29, Apr 7)&	2712 (1603, 3820)\\
$k-1$&    51.92 (42.08, 61.77)&	Apr 1 (Mar 28, Apr 6)&	2301 (1476, 3125)\\
$k-2$&	43.98 (36.76, 51.20)&	Mar 31 (Mar 27, Apr 5)&	1908 (1310, 2506)\\
$k-3$&	36.61 (30.96, 42.26)& Mar 30 (Mar 25, Apr 3)&	1531 (1114, 1949)\\
\hline
\end{tabular}
\end{table}

\begin{table}
\caption{\label{Parametes_Italy} \textbf{The parameter values and initial condition of simulations for Italy.} Phase 1 is between Jan 31 and Mar 8, and Phase 2 is between Mar 8 and Mar 26.}
\centering
\begin{tabular}{p{2.1cm}|p{2.1cm}|p{2.1cm}|p{3.2cm}|p{2.2cm}}
\hline
Parameter& Mean value & Standard derivation& $95\%$ CI &References \\
\hline
$\beta$ (Phase 1)&	0.0179&	$1.19\times10^{-3}$&	[0.0156, 0.0203]]&	MCMC\\
$\beta$ (Phase 2)&$4.45\times10^{-4}$&	$2.30\times 10^{-4}$&	$[0, 8.95\times 10^{-4}]$&	MCMC\\
$\xi$ (Phase 1)&	0.0996&	0.0458&	$[9.87\times 10^{-3}, 0.1894]$&	MCMC\\
$\xi$  (Phase 2)&	0.1312&	0.0250&	[0.0823, 0.1801]&	MCMC\\
$\sum_k N_k(0)$&59430000&	 -	&-	&\cite{WHOita2020}\\
$\sum_k S_k(0)$&	59429892	&-	&-&	Calculated\\
$\sum_k E_k(0)$&	69.01&	56.92&	[0, 180.58]&	MCMC\\
$\sum_k A_k(0)$&	22.60&	18.24&	[0, 58.35]&	MCMC\\
$\sum_k I_k(0)$&	32.11&	24.23&	[0, 79.59]&	MCMC\\
$\sum_k H_k(0)$&	2&	-&	-&	\cite{WHO_report2020}\\
$\sum_k R_k(0)$&	0&	-&	-&	Estimated\\
$\sum_k D_k(0)$&	0&	-&	-&	Estimated\\
\hline
\end{tabular}
\end{table}

\begin{table}
\caption{\label{Peak_Italy} \textbf{The peak number of new cases, peak time, and final size after containment strategies are implemented in Italy.}}
\centering
\begin{tabular}{p{1.5cm}|p{3.2cm}|p{3cm}|p{5.5cm}}
\hline
Scenarios &Peak size (95\%CI)&	Peak time (95\%CI)	&Final size (95\%CI)\\
\hline
$\beta$	&5492 (5277, 5708)&	Mar 26 (24, 27)&	$2.59\times 10^5 (2.10\times10^5, 3.09\times10^5)$\\
$0.8\beta$	&4340 (4235, 4564)&	Mar 25 (24, 26)&	$2.03\times 10^5 (1.75\times 10^5, 2.31\times10^5)$\\
$0.6\beta$&	3323 (3197, 3450)&	Mar 25 (24,  26)&	$1.52\times 10^5 (1.37\times 10^5, 1.66\times10^5)$\\
$q=0$&	5492 (5277, 5708)&	Mar 26 (24, 27)&	$2.59\times10^5 (2.10\times 10^5, 3.08\times10^5)$\\
$q=1/8$&	3413 (3279, 3547)&	Mar 20 (19, 21)&	$1.22\times10^5 (1.14\times10^5, 1.30\times10^5)$\\
$q=1/4$&	2609 (2434, 2783)&	Mar 18 (17, 19)&	$8.51\times10^4 (7.91\times10^4, 9.11\times10^4)$\\
\hline
\end{tabular}
\end{table}

\begin{table}
\caption{\label{VaringPeak_Italy} \textbf{The peak size, peak time and final  size for Italy when varying node degrees of
symptomatically infected individuals.}}
\centering
\begin{tabular}{p{1.2cm}|p{3.2cm}|p{2.9cm}|p{5.5cm}}
\hline
Degree &Peak size (95\%CI)&	Peak time (95\%CI)	&Final size (95\%CI)\\
\hline
$k$	&5492 (5277, 5708)&	Mar 26 (24, 27)	   &    $2.59\times 10^5 (2.10\times10^5, 3.08\times10^5)$\\
$k-1$&	4583 (4410, 4755)&	Mar 25 (24, 27)&	$2.13\times10^5 (1.81\times10^5, 2.44\times10^5)$\\
$k-2$&	3683 (3546, 3821)&	Mar 25 (24, 26)&	$1.69\times10^5 (1.50\times10^5, 1.88\times10^5)$\\
$k-3$&	2795 (2680, 2910)&	Mar 24 (23,  26)&	$1.28\times10^5 (1.17\times10^5, 1.39\times10^5)$\\
\hline
\end{tabular}
\end{table}

\section*{Acknowledgements}
LX is funded by  Fundamental Research Funds for the Central Universities of China. WS is funded by the National Science Foundation for Young Scholars of Heilongjiang Province QC2018004, and Fundamental Research Funds for the Central Universities of China (HEUCF181106, 3072019CF2411). JGEF is supported by multidisciplinary grant SIP-IPN 20196759. HZ is supported by Canadian Institutes of Health Research (CIHR), Canadian COVID-19 Math Modelling Task Force, and York Research Chair program of York University.
%
%\section*{Contributors}
%JGEF, JMH, JCM, WS, LX, and ZH conceived the study. SJ, JCM, WS, and LX designed and analysed the model. SJ and LX programmed the model and made the figures and tables. JGEF, HL, WS, and LX worked on statistical aspects of the study. All authors interpreted the results, contributed to writing the article, and approved the final version for submission.
%
\section*{Conflict of Interests Statement}
We declare that there is no conflict of interest associated
with this work.



%  \bibliographystyle{elsarticle-num}
% \bibliography{COVID20200502}


\end{document}